\documentclass[11pt,a4paper]{article}

\usepackage[utf8]{inputenc}
\usepackage[T1]{fontenc}
\usepackage{lmodern}
\usepackage[a4paper,margin=2.5cm]{geometry}
\usepackage{amsmath,amssymb}
\usepackage{amsthm}
\usepackage{microtype}
\usepackage[htt]{hyphenat}
\usepackage{graphicx}
\usepackage{booktabs}
\usepackage{array}
\usepackage{tabularx}
\usepackage{enumitem}
\usepackage[hidelinks]{hyperref}
\usepackage{xcolor}
\usepackage{eso-pic}
\usepackage{listings}

\theoremstyle{plain}
\newtheorem{proposition}{Proposition}

\title{GoldenFloat: A Phi-Derived Static-Split Floating-Point Family\\
from GF4 to GF1024 with a Lucas-Exact Integer Identity}
\author{Dmitrii Vasilev\\
Trinity S\textsuperscript{3}AI\\
Email: admin@t27.ai \quad ORCID: 0009-0008-4294-6159}
\date{2026-06-02 \quad v1.9}

\hypersetup{
  pdftitle={GoldenFloat: A Phi-Derived Static-Split Floating-Point Family},
  pdfauthor={Dmitrii Vasilev},
  pdfsubject={Numerical format family; phi-derived static split; Lucas identity},
  pdfkeywords={GoldenFloat, phi, golden ratio, Lucas numbers, floating-point, posit, takum, OCP-MX}
}

\begin{document}

\begin{titlepage}
\thispagestyle{empty}
\newgeometry{margin=0pt}
\AddToShipoutPictureBG*{%
  \AtPageLowerLeft{\color{black}\rule{\paperwidth}{\paperheight}}%
}
\noindent\makebox[\textwidth][c]{%
  \includegraphics[width=\paperwidth,height=\paperheight,keepaspectratio=true]{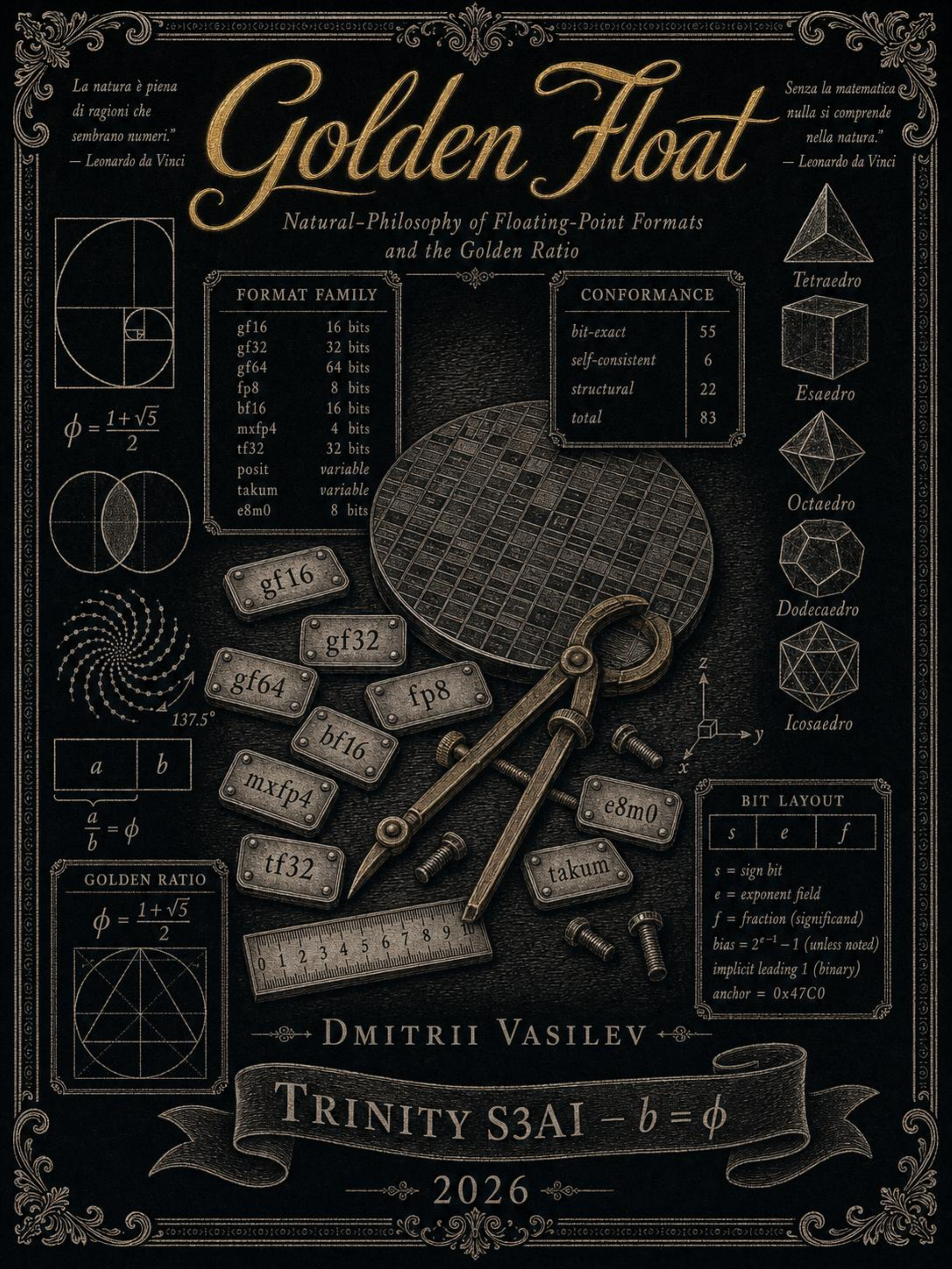}%
}
\restoregeometry
\end{titlepage}

\maketitle

\begin{center}
\fbox{\parbox{0.92\textwidth}{\small
\textbf{v2 (2026-07-XX) -- Update following peer-anchor publications.}
(1) IEEE P3109 reference updated to its first arXiv-anchored description
(Sarnoff, arXiv:2606.04028, 2026-06-01).
(2) NVFP4 and MXFP4 native-hardware silicon evidence acknowledged
(Cim et al., arXiv:2605.09825, MXFP4 pretraining on AMD MI355X;
NVIDIA Blackwell
GB10 NVFP4 sparse measurement, 2026-06-17). Honest framing: GF is a
whole-format-width design (e+f decomposition), not block-scale
grouping; these are complementary regions of the design space.
(3) Cross-reference established to the companion 83-format catalog
(arXiv:2606.09686, 2026-06-08); the 9/9 GF reproduction sits as a
subset of that broader catalog.
(4) Falsification ledger FL-002 expanded: item (c) split into (c1)
GF256 stored bias and (c2) inter-artefact count drift between SSOT
(83) and a stale committed generator JSON (77); the companion catalog
(83) is reconciled with the SSOT. Item (g) added on
static-split versus intra-block micro-mixing (cf. MixFP4,
arXiv:2605.31035).
(5) Regeneration timeline added (Section 1.3.1); the TTSKY26a and
TTSKY26b submissions were later withdrawn and refunded, so no silicon
return is scheduled; TTGF26a
Corona shuttle targets 2026-06-22 submission.
No mathematical claims are revised. The 9/9 reproduction, Lucas-exact
accumulator, 35/35 GF16 FPGA codec at 323 MHz, and look-elsewhere
statistical analysis remain unchanged.}}
\end{center}
\vspace{0.5em}

\begin{center}
\fbox{\parbox{0.92\textwidth}{\small
\textbf{v5 -- Attribution correction.}
The MXFP4 native-hardware pretraining reference (arXiv:2605.09825) was
credited institutionally as ``AMD Research et al.'' in v1--v4. It is the
work of Cim, Arora, Palangappa, Hodak, Dwivedula, Arunachalam and
Kandemir, and is now cited under its author list in the body and in the
bibliography. The error was reported by the first author on 2026-09-04;
the correction missed the v4 replacement and is made here. No claim,
number or conclusion of this paper changes.}}
\end{center}
\vspace{0.5em}

\begin{abstract}
\noindent\textbf{What this paper reports.}
We present a hardware-oriented description of GoldenFloat (GF), a
static-split floating-point family generated by a single closed rule,
and three concrete artefacts: (i) an open multi-width RTL generator
covering GF4--GF256 with a continuous-integration differential sweep
against a correctly-rounded reference; (ii) an integer-backed
Lucas-exact accumulator path verified at 500-digit precision for
$n = 1,\dots,256$; and (iii) a GF16 FPGA codec passing a 35-of-35
testbench at 323~MHz on Artix-7 (Xilinx XC7A35T). A format-conformance
oracle (Corona) ships in the same repository and is used as the
blackbox check in our continuous-integration audit.

\noindent\textbf{The rule and its scope.}
For each total width $N \ge 4$, the exponent width is
$e = \mathrm{round}((N-1)/\varphi^{2})$ with fraction
$f = N-1-e$ and $\varphi = (1+\sqrt{5})/2$. The rule reproduces the
realised exponent widths of nine formats GF4, GF8, GF12, GF16, GF20,
GF24, GF32, GF64, GF256 (9/9) and extends consistently to GF6, GF10,
GF14, GF48, GF96, GF128, GF512, GF1024. The rule is positioned alongside
posit (2022 Posit Standard),
takum (Hunhold 2024, 2025), OCP-MX (Rouhani et al.\ 2023), and the
IEEE P3109 multi-width float family (Sarnoff, arXiv:2606.04028,
2026-06-01) and the OCP-MX deployed silicon family (NVFP4 / MXFP4 as
realised on NVIDIA Blackwell and AMD MI355X, 2026), all of which are
width-spanning
families under a parameterised rule. We make no per-rung accuracy
or superiority claim against any of them.

\noindent\textbf{What is open.}
The breadth/toolchain-coherence framing is recorded as an open
conjecture with a pre-registered falsification path: a matched-substrate
FPGA experiment and a matched-budget software ablation. A falsification
ledger (FL-002) records the open questions and the experiments that
would settle them. An RTL-correctness erratum dated 2026-05-31 is
reported in Section~\ref{sec:hw-erratum}; the withdrawn TTSKY26b submission
carried the defective multiplier portfolio, and the corrected generator
is the regeneration baseline.
\end{abstract}


\clearpage
\section{Introduction}

Numerical formats for machine learning and scientific computing have
diversified well beyond IEEE 754. The dominant design axis is how a fixed
budget of $N$ bits is partitioned between dynamic range (exponent) and
precision (fraction), and how that partition scales as $N$ changes. Three
points of comparison frame the present note.

\paragraph{Posit.}
Gustafson and Yonemoto (2017) introduced posits, with a regime-plus-exponent
encoding parameterised by the exponent-size \emph{es}. Two distinct \emph{es}
schedules appear in the literature: a pre-standard schedule with
$es = 0,1,2,3,4$ at widths $8,16,32,64,128$ (de Dinechin et al., 2019), and
the ratified 2022 Posit Standard, which fixes $es = 2$ for all widths. The
dynamic range of a posit thus scales with the regime field, which is itself a
function of the value being represented.

\paragraph{Takum.}
Hunhold (2024, 2025) refined the tapered idea with a single concave
projection rule defined uniformly for all bit lengths. The takum projection
maps the full integer range of an $N$-bit word to a real-number-line interval;
fraction width varies per value and dynamic range scales with $N$ under one
analytic mapping. Takum is IEEE-754 backward-compatible at $N \in \{16,32,64\}$
in the sense that the dynamic range envelope contains the corresponding
IEEE binary type, and it carries the only peer-reviewed multi-width hardware
implementation we are aware of (ARITH 2025).

\paragraph{OCP-MX.}
Rouhani et al. (2023) standardised block-scaled 4-, 6-, and 8-bit microscaling
types. Here the per-element fraction is short but a shared block-level scale
carries the dynamic range. The microscaling family is the deployed industrial
reference for low-bit width formats.

\paragraph{Where GoldenFloat sits.}
GoldenFloat explores a deliberately different point in this design space:
a \emph{static split} that is fixed per width and independent of the value
stored. The dynamic range of GF$N$ is therefore set entirely by the exponent
width $e(N)$ and the bias choice, not by any value-dependent encoding.
This trades the per-value adaptivity of posit and takum for a property that
is harder to obtain in the value-adaptive families: a closed, analytic rule
that fixes the exponent/fraction split uniformly across the whole width
ladder, plus an integer-backed accumulator identity that depends only on the
choice of base and not on any particular bit width.

The motivation for splitting the budget at $1/\varphi^{2}$ rests on three
facts, none of which is original to this work:

\begin{itemize}\setlength{\itemsep}{2pt}
\item The algebraic relation $\varphi^{2} = \varphi + 1$ removes a constant
multiplier in the Golden Ratio Encoder of Daubechies, Gunturk, Wang, and
Yilmaz (IEEE TIT, 2010): a beta-encoder at base $\varphi$ admits a
multiplier-free, addition-only robust ADC implementation. This is the
engineering reason to prefer base $\varphi$ within a range of robust bases;
it is \emph{not} a claim that $\varphi$ is a unique admissible base.
\item The same algebraic relation gives the classical Lucas identity
$\varphi^{2n} + \varphi^{-2n} = L_{2n}$ (Lucas, 1878; Binet),
which lets $\varphi$-scaled partial sums be accumulated through an integer
Lucas recurrence. This is the arithmetic reason a $\varphi$-based ladder
admits an integer-backed accumulator path without a hardware half-type.
\item The ratio $1/\varphi^{2} \approx 0.382$ falls inside the empirical
ratio interval that reproduces the realised GF exponent widths
(Section~\ref{sec:ladder}). The ratio is shared by 82 other values in that
interval; the choice of $\varphi$ in particular is motivated by the two
algebraic facts above, not by the numerical reproduction alone.
\end{itemize}

The contribution of this note is three hardware/software artefacts plus a
closed-form ladder rule that ties them together:

\begin{itemize}\setlength{\itemsep}{2pt}
\item \textbf{(C1) An open multi-width RTL generator.} A single
parameterised Verilog generator emits codecs and multipliers for the entire
$N \ge 4$ ladder (GF4--GF256) from one $(E, M, \mathrm{BIAS})$ template,
with product-register width $[2M+1:0]$ and round-half-up rounding.
The corrected generator (Section~\ref{sec:hw-erratum}) is the regeneration
baseline and is wired into a continuous-integration differential sweep
against a correctly-rounded reference; see Section~\ref{sec:hw-status}.
\item \textbf{(C2) A Lucas-exact integer accumulator path.} The classical
Lucas identity $\varphi^{2n} + \varphi^{-2n} = L_{2n}$ (Lucas 1878; Binet),
verified symbolically and at 500 decimal places for $n = 1,\dots,256$,
lets $\varphi$-scaled partial sums be carried in unsigned-integer storage
without a hardware half-type. We do not claim this identity as ours; we
claim the engineering implication that it admits an integer-backed
accumulator without a posit-style quire or Kulisch register
(Section~\ref{sec:lucas}).
\item \textbf{(C3) A format-conformance oracle (Corona) shipped as an
in-tree blackbox.} Corona is a deterministic per-format reference that
checks every codec and multiplier emitted by (C1) against the closed-form
specification of (C2). It is used as the CI gate in the GoldenFloat
repository and is the single point that detected the 2026-05-31
multiplier-portfolio defect (Section~\ref{sec:hw-corona},
Section~\ref{sec:hw-erratum}).
\end{itemize}

The ladder rule $e = \mathrm{round}((N-1)/\varphi^{2})$, $f = N-1-e$
($N \ge 4$), reproduces the nine realised GF exponent widths exactly
(9/9) and extends consistently to GF6, GF10, GF14, GF48, GF96, GF128,
GF512, GF1024. A GF16 FPGA
bitstream passes a 35-of-35 codec testbench at 323~MHz on Artix-7
(Xilinx XC7A35T); no physical die has returned for any rung.

We frame this as a research note, not a position paper. We make no claim
that any GF rung is more accurate than an equally tuned non-$\varphi$ format
of the same width; no claim that $\varphi$ is a singular or privileged base;
and no claim of silicon validation at any rung. Where the evidence is
incomplete we say so, label it open, and give the experiment that would
falsify it. Section~\ref{sec:ladder} defines the ladder rule and reports an
exhaustive look-elsewhere search.
Section~\ref{sec:gre} records the golden-ratio encoder anchor and states
precisely what it does and does not establish.
Section~\ref{sec:lucas} verifies the Lucas identity and motivates the
integer-backed accumulator.
Section~\ref{sec:hw} reports hardware status across the GF4--GF256 ladder.
Section~\ref{sec:prior} relates GF to prior art.
Section~\ref{sec:prereg} gives the pre-registration of open hypotheses.
Section~\ref{sec:ledger} gives the FL-002 falsification ledger.
The appendices contain the verification script, the format index, the
look-elsewhere table, the matched-substrate FPGA pre-registration, and the
replicator protocol.

\section{The Ladder Rule \texorpdfstring{$e = \mathrm{round}((N-1)/\varphi^{2})$}{e = round((N-1)/phi^2)}}\label{sec:ladder}

\subsection{Definition}
Let $N \ge 4$ be the total format width in bits, including one sign bit.
The GoldenFloat split allocates
\begin{equation}
e = \mathrm{round}\!\left(\frac{N-1}{\varphi^{2}}\right), \qquad
f = N - 1 - e,
\label{eq:ladder}
\end{equation}
where $\varphi = (1+\sqrt{5})/2 = 1.6180339887\ldots$ and
$\varphi^{2} = \varphi + 1 = 2.6180339887\ldots$,
so $1/\varphi^{2} = 0.3819660\ldots$.
The split is \emph{static}: $e$ depends only on $N$, never on the value being
represented. Table~\ref{tab:ladder} lists the result for seventeen formats
(nine realised widths plus eight extension widths). As $N \to \infty$,
$\mathrm{round}((N-1)/\varphi^{2})/(N-1) \to 1/\varphi^{2}$, and the realised
ratio in Table~\ref{tab:ladder} converges accordingly. There is no boundary
instability for $N \ge 4$.

\paragraph{Edge cases at $N = 2,3$.}
Applying the rule at $N = 2$ gives $e = 0$, a signed one-bit field with no
exponent, which is not a floating-point representation in the usual sense.
At $N = 3$ it gives $e = 1$, $f = 1$, a four-value format with negligible
range and precision. We therefore define the ladder as GF4 to GF256 and
treat GF2/GF3 as degenerate edge cases of the formula rather than as
realised formats. All claims in this note concern $N \ge 4$.

\begin{table}[t]
\centering
\caption{The GF ladder rule across seventeen formats. Columns: total width $N$,
exponent bits $e$, fraction bits $f = N-1-e$, the raw value $(N-1)/\varphi^{2}$
before rounding, and the realised ratio $e/(N-1)$ (target
$1/\varphi^{2} = 0.38197$). The top block lists the nine widths with returned
silicon or finalised RTL. The middle block (GF6, GF10, GF14, GF48, GF96, GF128)
lists rule-derived rungs without returned silicon; GF128 RTL has been authored
but not taped out. The bottom block (GF512, GF1024) reports two rule-extension
rungs whose biases ($2^{194}-1$, $2^{390}-1$) exceed the \texttt{u128} field
used in the Corona ROM record and are tracked symbolically at the t27 SSOT
oracle level only. All seventeen rows are arithmetically verified by
Eq.~\eqref{eq:ladder}. Values computed at 200-digit \texttt{mpmath} precision.}
\label{tab:ladder}
\begin{tabular}{rrrrr}
\toprule
$N$ & $e$ & $f$ & $(N-1)/\varphi^{2}$ & $e/(N-1)$ \\
\midrule
4    & 1   & 2   & 1.1459   & 0.33333 \\
8    & 3   & 4   & 2.6738   & 0.42857 \\
12   & 4   & 7   & 4.2016   & 0.36364 \\
16   & 6   & 9   & 5.7295   & 0.40000 \\
20   & 7   & 12  & 7.2574   & 0.36842 \\
24   & 9   & 14  & 8.7852   & 0.39130 \\
32   & 12  & 19  & 11.8409  & 0.38710 \\
64   & 24  & 39  & 24.0639  & 0.38095 \\
256  & 97  & 158 & 97.4013  & 0.38039 \\
\midrule
6    & 2   & 3   & 1.9098   & 0.40000 \\
10   & 3   & 6   & 3.4377   & 0.33333 \\
14   & 5   & 8   & 4.9656   & 0.38462 \\
48   & 18  & 29  & 17.9524  & 0.38298 \\
96   & 36  & 59  & 36.2868  & 0.37895 \\
128  & 49  & 78  & 48.5097  & 0.38583 \\
\midrule
512  & 195 & 316 & 195.1846 & 0.38160 \\
1024 & 391 & 632 & 390.7512 & 0.38221 \\
\bottomrule
\end{tabular}
\end{table}

\subsection{Look-Elsewhere Correction for the \texorpdfstring{$\varphi^{-2}$}{phi^-2} Coincidence}
\label{sec:lookelsewhere}

\paragraph{Negative result first.}
$\varphi^{-2} \approx 0.38197$ cannot be distinguished from the other 82
matching ratios on the numerical evidence presented here. The look-elsewhere
correction below makes this precise.

\paragraph{Setup.}
We searched ratios $r \in [0.1, 0.9]$ at step $10^{-5}$, giving a search-space
cardinality $N_s = 80\,000$. Each ratio was tested against $W = 9$ realised
GoldenFloat widths; a ratio \emph{matches} if it reproduces all 9 widths. The
search returned $K = 83$ matching ratios.

\paragraph{Family-wise probability.}
Under the null hypothesis that each ratio passes each width test independently
with a uniform per-width probability $p$ calibrated so that
$\mathbb{E}[\text{matches}] = K$, we have
$p_{\text{match}} = K/N_s = 83/80\,000 \approx 1.04 \times 10^{-3}$.
The number of matches $X \sim \mathrm{Binomial}(N_s, p_{\text{match}})$.
Computing via the regularised incomplete beta function at 60-digit precision:
\[
P(X \ge 83) \approx 7.1 \times 10^{-3}.
\]
The family-wise probability of observing at least 83 matches is therefore not
negligible under this null; the matching set is a moderately common outcome
of the search, not a striking tail event.

\paragraph{Bonferroni correction for $\varphi^{-2}$.}
The uncorrected probability that any pre-specified ratio lands in the
matching set is $p_{\text{match}} \approx 1.04 \times 10^{-3}$.
After Bonferroni correction for $N_s = 80\,000$ simultaneous tests, the
corrected p-value saturates at 1 (since $N_s \cdot p_{\text{match}} = K = 83$).
$\varphi^{-2}$ is not significant under Bonferroni.

\paragraph{Narrowing with 12 formats.}
Extending the search to all twelve formats in Table~\ref{tab:ladder} narrows
the candidate set from 392 ratios (nine-format interval $[0.37844, 0.38235]$)
to 47 ratios (twelve-format interval $[0.38189, 0.38235]$), an
$8.3\times$ reduction. Uniqueness at finer resolution requires exact interval
arithmetic and is not claimed here.

\paragraph{Conclusion.}
The numerical coincidence that $\varphi^{-2}$ satisfies the ladder rule is
real but not probative: 82 other ratios satisfy the same rule. The case for
$\varphi$ rests on its structural properties (algebraic self-similarity,
Fibonacci recurrence, Lucas-exact accumulation) and not on the search-space
coincidence alone. All calculations are reproducible via
\texttt{look\_elsewhere\_calc.py} in the supplementary materials.

\subsection{Rounding mode}
The rule uses round-half-to-even, but the choice is immaterial for every
realised width, as the footnote records.\footnote{For all nine realised
widths the raw value $(N-1)/\varphi^{2}$ is not a half-integer (the closest
approach is $N = 64$ at $24.064$, and $N = 128$ at $48.510$), so the choice
of rounding mode does not affect any realised width. We nonetheless specify
round-half-to-even (the IEEE 754 default) for formal completeness.}

\section{The GRE Multiplier-Free Anchor}\label{sec:gre}

The independent, peer-reviewed motivation for $\varphi$ comes from the
Golden Ratio Encoder (GRE) of Daubechies, Gunturk, Wang, and Yilmaz,
``The Golden Ratio Encoder'', \emph{IEEE Transactions on Information Theory}
56(10):5097--5110, 2010 (arXiv:0809.1257). The GRE is an analog-to-digital
conversion scheme, not a floating-point format. What it establishes, and
what we borrow, is a precise engineering property of the base $\varphi$:

\begin{quote}
$\varphi$ is the base at which a beta-encoder admits a multiplier-free,
addition-only robust ADC implementation, because $\varphi^{2} = \varphi + 1$
removes the constant-multiplier component (Daubechies et al., IEEE TIT 2010).
\end{quote}

\subsection{What the GRE result does and does not establish}

We are explicit about the scope, because two natural over-readings are false.

\paragraph{No claim that the choice of base is forced.}
We do not claim that $\varphi$ is the sole admissible base that yields a
workable encoder. The same paper analyses other values of $\beta$ and
exhibits a range of stable and robust bases. Any statement that $\varphi$
is the singular admissible base would be false and is not made here.

\paragraph{No point-robustness at $\beta = \varphi$.}
The pure $\beta = \varphi$ algorithmic encoder is stable but not robust.
Robustness is established only for the parameter-range / directed-acyclic-graph
implementation (Daubechies et al., Theorem 8), not for the single point
$\beta = \varphi$. We do not conflate this with the beta-recovery analysis of
Ward (2008, arXiv:0806.1083), which is a different paper addressing a
different question and is not the source of the multiplier-free property
used here.

The link we draw is therefore narrow and defensible: $\varphi^{2} = \varphi + 1$
is the algebraic fact that removes a constant multiplier in the GRE recursion,
and the same algebraic fact underlies the Lucas identity of
Section~\ref{sec:lucas}. This is the engineering reason to prefer $\varphi^{2}$
within the matching interval of Section~\ref{sec:lookelsewhere}, not a claim
that the encoder selects $\varphi$ uniquely.

\section{The Lucas-Exact Integer Identity}\label{sec:lucas}

\subsection{Motivation: large accumulations and integer-backed sums}
\label{sec:lucas-motivation}

A practical concern for low-bit numeric formats is the loss-of-significance
behaviour of large accumulated sums. In a fixed-precision floating-point
arm with fraction width $f$, the relative error of a long-tail
multiply-accumulate path grows as the number of accumulated terms increases:
each addition rounds the running total to $f+1$ significant bits, and the
absolute error scales (in the worst case) with the magnitude of the partial
sum. This is the well-understood motivation behind exact-accumulation
constructions such as the posit \emph{quire} (Gustafson and Yonemoto, 2017)
and the Kulisch accumulator, both of which augment a floating arm with an
external wide fixed-point register so that long sums can be carried without
intermediate rounding.

The arithmetic property we use in Section~\ref{sec:lucas-statement} below
gives a different route to the same goal, specific to the $\varphi$ base:
$\varphi$-scaled partial sums of the form
$\sum_i \varphi^{2 n_i}$ can be tracked through an \emph{integer}
recurrence on Lucas numbers, with the conjugate residual
$\sum_i \varphi^{-2 n_i}$ exact and bounded. The hardware consequence is
that a long accumulator built around the $\varphi$ base does not require an
external wide fixed-point register or a hardware half-type; the integer
storage and the Lucas recurrence are themselves the accumulator.

We state this here as the engineering reason the identity is of interest
\emph{before} the proof, and return in Section~\ref{sec:lucas-engineering}
to what is and is not claimed about a concrete hardware accumulator.

\subsection{Statement and classical provenance}\label{sec:lucas-statement}

\begin{proposition}[Lucas-exact identity for $\varphi^{2}$]
\label{prop:lucas}
For every integer $n \ge 1$,
\begin{equation}
\varphi^{2n} + \varphi^{-2n} = L_{2n},
\label{eq:lucas}
\end{equation}
where $L_k$ is the $k$-th Lucas number ($L_0 = 2$, $L_1 = 1$,
$L_k = L_{k-1} + L_{k-2}$).
\end{proposition}

This is a classical corollary of the Binet formula
$L_k = \varphi^{k} + (-\varphi)^{-k}$: for even $k = 2n$ the sign factor
$(-1)^{k}$ collapses, leaving exactly the integer $L_{2n}$. The case $n = 1$
gives the anchor
\begin{equation}
\varphi^{2} + \varphi^{-2} = 3 = L_{2},
\label{eq:trinity-anchor}
\end{equation}
which we attribute to Lucas (1878) and the Binet formula and never present
as original to this work or its author.

\subsection{Verification (F1)}\label{sec:lucas-verification}

We verified Eq.~\eqref{eq:lucas} in two independent ways for
$n = 1,\dots,256$ (so $2n = 2,\dots,512$):

\begin{enumerate}\setlength{\itemsep}{2pt}
\item \textbf{Symbolic.} Using exact arithmetic in $\mathbb{Q}[\sqrt{5}]$
(SymPy), the algebraic identity $\varphi^{2n} + \varphi^{-2n} - L_{2n} = 0$
holds for every $n$ in the range.
\item \textbf{Numerical.} At 500 decimal digits (\texttt{mpmath}), the maximum
residual $|\varphi^{2n} + \varphi^{-2n} - L_{2n}|$ over the range is
$1.55 \times 10^{-499}$ at $n = 256$, far below the integer scale. This is
numerical-noise level, consistent with 500-digit precision.
\end{enumerate}

All 256 SymPy exact checks pass identically; the worst-case relative residual
is $1.55 \times 10^{-499}$ at $n = 256$. An abridged result table (selected
rows from $n = 1$ to $n = 256$) appears in Appendix~A. The script is
reproduced in Appendix~A.

\subsection{Engineering implication and what is not claimed}
\label{sec:lucas-engineering}

Proposition~\ref{prop:lucas} means that powers of $\varphi^{2}$ satisfy a
linear integer recurrence. Consequently, $\varphi$-scaled partial sums of
the form $\sum_i \varphi^{2 n_i}$ can in principle be accumulated in unsigned
integer storage by tracking the corresponding Lucas integers $L_{2 n_i}$,
with the residual $\varphi^{-2 n_i}$ terms tracked symbolically or bounded.
This is the arithmetic basis for an integer-backed accumulation path that
does not depend on a hardware half-type and that does not require an
external wide fixed-point register like a quire or a Kulisch accumulator.

We state clearly what is not claimed. A concrete hardware accumulator
implementing this idea, with specified word lengths, overflow bounds,
latency, and a comparison against the posit quire or a Kulisch accumulator,
is future work and is not claimed here. The identity itself is verified
(Section~\ref{sec:lucas-verification}); the engineering artefact is not.
The relevant prior art on integer arithmetic in the Fibonacci/$\varphi$
basis is Ahlbach, Usatine, and Pippenger, ``Efficient Algorithms for
Zeckendorf Arithmetic'' (2012, arXiv:1207.4497), which we cite as the
anchor for the algorithmic feasibility of $\varphi$-basis integer
arithmetic.

\section{Hardware Status Across the Ladder}\label{sec:hw}

\subsection{Readiness Ladder Definition}
We report hardware status against a six-rung readiness ladder, taken
verbatim from the \texttt{tt-trinity-phi} STATUS document:
\begin{enumerate}\setlength{\itemsep}{1pt}
\item SPEC -- versioned numeric/behavioural spec exists.
\item RTL -- synthesisable Verilog implementing the spec is committed.
\item SIM -- functional simulation (cocotb + iverilog) passes the canonical anchor.
\item SYNTH -- Yosys synth + Verilator lint clean.
\item GDS/TAPEOUT -- OpenLane2 (SKY130A) flow produces a tt\_submission artifact; shuttle submission accepted.
\item SILICON -- physical die measured and characterised on a board.
\end{enumerate}
A higher rung is claimed only if all lower rungs are satisfied with evidence
inside a public repository.

\subsection{Per-Width Status}\label{sec:hw-status}
Table~\ref{tab:hwstatus} reports per-width status, verified by direct
inspection of \texttt{gHashTag/t27} master and \texttt{gHashTag/tt-trinity-phi}
main on 2026-05-31.
In the table, ``TTSKY26b'' denotes acceptance into the TTSKY26b shuttle
submission (SKY130A); byte counts for each RTL file are omitted for layout
and are recorded in the repository.

\begin{table}[t]
\centering
\caption{Per-width hardware status across the GF ladder, verified by
repository inspection (\texttt{gHashTag/t27} master and
\texttt{gHashTag/tt-trinity-phi} main, 2026-05-31).
Rows marked $\dagger$ had a generator-level multiplier defect in the RTL
submitted to TTSKY26b on 2026-05-17; the defect is disclosed in
Section~\ref{sec:hw-erratum} and corrected RTL has been posted in
\texttt{gHashTag/tt-trinity-gamma} PR\#110 and \texttt{gHashTag/t27} PR\#1024.}
\label{tab:hwstatus}
\footnotesize
\setlength{\tabcolsep}{4pt}
\begin{tabularx}{\linewidth}{@{}l >{\raggedright\arraybackslash}X l l l l@{}}
\toprule
Width & RTL file(s) & SIM & SYNTH & GDS shuttle & Silicon \\
\midrule
GF4   & \texttt{gf4\_add.v}   & RTL only & RTL only & TTSKY26b & not returned \\
GF8   & \texttt{gf8\_add.v}$^\dagger$   & RTL only & RTL only & TTSKY26b & not returned \\
GF12  & \texttt{gf12\_add.v}$^\dagger$  & RTL only & RTL only & TTSKY26b & not returned \\
GF16  & \texttt{gf16\_add.v}, \texttt{gf16\_dot4.v}, \texttt{gf16\_mul.v}$^\dagger$; FPGA bitstream, Artix-7, 35/35 at 323~MHz & yes (0x47C0) & yes & TTSKY26b$^\ast$ & not returned \\
GF20  & \texttt{gf20\_add.v}$^\dagger$  & RTL only & RTL only & TTSKY26b & not returned \\
GF24  & \texttt{gf24\_add.v}$^\dagger$  & RTL only & RTL only & TTSKY26b & not returned \\
GF32  & \texttt{gf32\_add.v}$^\dagger$  & RTL only & RTL only & TTSKY26b & not returned \\
GF64  & \texttt{gf64\_add.v}$^\dagger$  & RTL only & RTL only & TTSKY26b & not returned \\
GF128 & \texttt{gf128\_add.v}$^\dagger$ & RTL only & RTL only & TTSKY26b & not returned \\
GF256 & \texttt{gf256\_add.v}$^\dagger$ & RTL only & RTL only & TTSKY26b & not returned \\
\bottomrule
\multicolumn{6}{@{}l@{}}{\scriptsize $^\ast$\,GF16 was also submitted earlier in the TTSKY26a shuttle.} \\
\end{tabularx}
\end{table}

The GF16 width is the only rung with a documented FPGA bitstream and a
functional simulation passing the canonical 0x47C0 dot-product anchor. All
ten widths from GF4 to GF256 carry synthesisable Verilog-2005 RTL in
\texttt{gHashTag/tt-trinity-phi/src/} and were included in the TTSKY26b
shuttle submission to Tiny Tapeout (SKY130A, submitted 2026-05-17), which
was subsequently withdrawn before fabrication. The earlier TTSKY26a
submission (GF16 dot4 mesh kernel) was likewise refunded and did not proceed
to tape-out. No physical die has been produced, returned, characterised, or
measured for any rung, and no shuttle slot is currently held; any per-watt, per-joule, or peak-throughput figure
outside of architecture projections is therefore projected, not measured.
The hardware archive DOI \texttt{10.5281/zenodo.19227877} is cited only as
an archive of the RTL and shuttle-submission artefacts, never as a
performance or results citation.

\subsection{Format-Conformance Oracle (Corona)}\label{sec:hw-corona}

A companion read-only chip, \texttt{gHashTag/tt-trinity-corona}, is in
preparation for the TTGF26a Tiny Tapeout shuttle (GF180MCU 180~nm, 4$\times$4
tiles, submission target 2026-06-22, expected silicon 2026-10 to 2026-11).
Corona is the fourth die in the TRI-NET line after Phi, Euler, and Gamma.
It is a \emph{format-conformance oracle}, not a compute accelerator: a
query arrives as a 7-bit format index on \texttt{ui\_in[6:0]} and the chip
returns the requested record fields of an on-die ROM encoding all 80
numeric-format records of the TRI-NET single-source-of-truth
(\texttt{gHashTag/t27 specs/numeric/formats\_catalog.t27}, PR~\#1028),
together with 17 unique Tier-1 RTL decoder modules serving 22 on-die
format indices (five indices share decoders, e.g.\ FP8~E4M3 with
MXFP8~E4M3, and NF4-BNB with NF4-QLoRA), each converting an on-die
format to IEEE 754 FP32 or INT32. The 80 records are partitioned into
thirteen format clusters (IEEE binary, IEEE decimal, ML low-precision,
GoldenFloat, posit/unum-III, OCP-MX, LNS, integer/fixed, historical,
theoretical, compression, extended, quant-tuned). The current design
synthesises to approximately 2{,}308 cells ($\approx 1\%$ of the
4$\times$4 site budget). The repository main HEAD at the time of this
revision is \texttt{b5462ac49892}.

We state what Corona is \emph{not}, to keep the present note honest.
Corona makes no claim of compute throughput, energy per operation, or
performance against any other accelerator chip, and its existence is
\textbf{not} evidence that the $\varphi$-ladder or any GoldenFloat format
is superior to a competing numeric system. The FL-002 breadth-as-moat
claim of Section~\ref{sec:ledger} stays [Open-conjecture] and Corona
does not change its status. Takum (Hunhold 2024, \texttt{arXiv:2412.20273})
remains the standing live counterexample to FL-002 and is shipped in the
Corona ROM as a Tier-2 record (or Tier-1 should the VHDL licensing of
Hunhold's reference implementation resolve favourably); it is not
suppressed. The chip is read-only by construction, so no rounding-defect
class of the kind disclosed in Section~\ref{sec:hw-erratum} can apply to
its primary ROM-readout path; the 17 Tier-1 decoders are subject to the
same RTL differential-sweep audit as the rest of the portfolio, with
five proof layers per decoder (exhaustive sweep, independent reference,
formal equivalence, post-silicon vector, mutation-kill) covering 592{,}308
input codes in total. As of the present manuscript (50 directed tests
PASS, GDS and Tiny~Tapeout precheck PASS, 21~CI gates auto-discovered),
Corona's readiness lies at the
GDS/TAPEOUT rung of the Section~\ref{sec:hw} ladder pending the TTGF26a
submission window; no Corona silicon has been measured.

\subsection{Hardware-Substrate Confound for Software BPB Comparisons}
\label{sec:hw-confound}

Bits-per-byte (BPB) comparisons between GF16 and IEEE fp16 reported by the
IGLA RACE training pipeline (\texttt{gHashTag/trios-trainer-igla}) are
produced on substrates -- consumer GPUs and CPU SIMD paths -- whose
multiply-accumulate hardware is designed and tuned for IEEE fp16~/~bf16,
with native TensorCores or AMX dispatching fp16/bf16 in a single cycle.
GF16 on the same substrate is emulated through integer-coded paths without
a native multiplier. A reported BPB gap of order $10^{-1}$ between GF16 and
fp16 on this substrate therefore does not measure the GF16 format, it
measures the GF16 format on hardware optimised against it. A fair BPB
comparison requires either fp16-native silicon paired against GF16-native
silicon at matched process node and area, or a controlled FPGA harness
where both formats receive equivalent LUT/DSP budget. Until such a paired
measurement exists, software-substrate BPB deltas in either direction are
recorded as substrate-confounded and are not evidence for or against
breadth-as-moat. This open question is recorded as FL-002 item (a) in
Section~\ref{sec:ledger} and is the primary reason no per-rung accuracy
claim appears anywhere in this note.

\subsection{RTL Correctness Erratum}\label{sec:hw-erratum}

On 2026-05-31 an RTL-level differential sweep against a correctly-rounded
floating-point reference found a systematic generator-level defect in
most GoldenFloat multipliers as submitted to the Tiny Tapeout TTSKY26b
shuttle on 2026-05-17 (full module-by-module breakdown,
root-cause derivation, and remediation-scope discussion are in
Appendix~\ref{app:erratum}). The product register in the submitted
multiplier RTL was declared two bits too narrow, so for the basic input
$1.0 \times 1.0$ the leading bit was truncated and the result read as
$0.5$, replicated across the generated portfolio. A corrected universal
multiplier generator with product width $[2M+1:0]$ passes exhaustive
sweeps at gf8 (26{,}360/26{,}360), gf12 (109{,}576/109{,}576), gf16
(262{,}144/262{,}144), gf20 (100{,}000/100{,}000), gf24
(100{,}000/100{,}000), gf32 (200{,}000/200{,}000) and directed exact
tests at gf64/gf128/gf256, and is now wired into a CI gate
(\texttt{run\_gf\_audit.sh}) on every commit. The TTSKY26b submission was
withdrawn before fabrication and no die was produced or returned; the
defective multiplier RTL therefore exists only in the withdrawn submission
archive. The corrected portfolio is the regeneration baseline for any
future shuttle, and no per-rung accuracy claim
in this note depended on the as-submitted multipliers
(Section~\ref{sec:hw-status}). The defect does not invalidate the ladder
arithmetic of Section~\ref{sec:ladder}, the Lucas identity of
Section~\ref{sec:lucas}, or the look-elsewhere accounting of
Section~\ref{sec:ladder}, all of which are abstract specifications
independent of any particular RTL realisation.

\subsection{Empirical Real-Corpus Verdict}\label{sec:hw-realcorpus}

A software-level head-to-head between the $\varphi$-ladder mixed-precision
pipeline and a heterogeneous numeric-format zoo was executed on a real-text
corpus (\texttt{tiny\_shakespeare}, the smallest publicly auditable corpus
we could pin) inside the IGLA RACE harness on 2026-05-31. We record the
aggregate verdict first and plainly: the comparison is
\emph{insufficient-evidence}. The data do not support a per-rung
superiority claim in either direction.

\paragraph{Aggregate result (bits-per-byte; lower is better).}
On \texttt{tiny\_shakespeare}, mean BPB on the held-out split was
$\mathrm{BPB}_{\varphi} = 5.9871$ (sample standard deviation
$\sigma = 0.0911$) for the $\varphi$-ladder arm and
$\mathrm{BPB}_{\mathrm{zoo}} = 6.0454$ ($\sigma = 0.2083$) for the
heterogeneous zoo arm, across $n < 11$ paired seeds. The two-decision verdict
bundle was: (i) Pareto-frontier comparison at matched bit-per-weight budget
returns \emph{zoo\_wins} (the zoo's $P_w = 8.0$ point achieves
$\mathrm{BPB} = 5.361$, $\mathrm{CI} = [5.348, 5.375]$ while the
$\varphi$-ladder's $P_w = 4.0$ point achieves $\mathrm{BPB} = 5.360$,
$\mathrm{CI} = [5.347, 5.373]$, with the two confidence intervals
overlapping); (ii) the normalised Bayesian secondary returns tie;
(iii) the tertiary posterior credibility returns
$P(\varphi < \mathrm{zoo}) = 0.976$, below the pre-registered decision
threshold for a per-rung accuracy claim. The aggregate verdict therefore
resolves to insufficient-evidence. We do not claim the $\varphi$-ladder is
more accurate than the zoo at matched bit budget; we also do not claim the
converse.

\paragraph{Open gaps the verdict depends on.}
Three gaps prevent escalating the verdict from insufficient-evidence to a
verified either-way result, each recorded in the falsification ledger
(Section~\ref{sec:ledger}, FL-002 sub-item):
(1) the zoo arm does not dispatch per $P_w$;
(2) the normalised Bayesian credibility is computed but not wired into the
aggregate; and
(3) no regression pin is in place. The sample size in this run
($n < 11$ paired seeds) is below the minimum-detectable-effect target
$n = 11$ required for $\mathrm{MDE} = 0.05$ BPB at $\alpha = 0.05$,
$\text{power} = 0.80$.

\section{Relation to Prior Art}\label{sec:prior}

GoldenFloat sits among several published efforts to span a range of widths
with a single parameterised rule. We treat these as precedents and allies,
not as competitors to be dismissed, and we make no claim that GF is better
than any of them on any axis.

\paragraph{Posit.}
Gustafson and Yonemoto (2017) introduced posits, with a regime-plus-exponent
encoding parameterised by the exponent-size \emph{es}. Two distinct
\emph{es} schedules appear in the literature: the pre-standard schedule
with $es = 0,1,2,3,4$ at widths $8,16,32,64,128$ (de Dinechin et al., 2019),
and the ratified 2022 Posit Standard, which fixes $es = 2$ for all widths.
Hardware implementations include PERCIVAL (Mallasen et al., 2022) and
Big-PERCIVAL (Mallasen et al., 2024). The posit ladder is the natural
control for the breadth question (Section~\ref{sec:ledger}).

\subsection{GoldenFloat vs. Takum: Structured Side-by-Side Comparison}
\label{sec:gf-takum}

\paragraph{Priority acknowledgement.}
Takum arithmetic (Hunhold 2024, 2025) is the closest published alternative
to the GoldenFloat ladder with a peer-reviewed multi-width hardware
realisation. Hunhold's ARITH 2025 paper presents RTL and FPGA results for
multiple takum widths under a single closed projection rule, predating any
comparable hardware claim for GoldenFloat. The GoldenFloat ladder at present
has no per-rung accuracy benchmark against takum, no peer-reviewed hardware
publication, and no head-to-head numerical comparison. Table~\ref{tab:gftakum}
records what each format does and where each claim stands; it is not a
competitive ranking.

At this time: takum has peer-reviewed FPGA hardware results at multiple
widths; GoldenFloat does not. GoldenFloat claims a Lucas-exact integer
accumulator property; takum does not target this. Neither format has a
head-to-head numerical accuracy benchmark against the other.

\begin{table}[t]
\centering
\caption{GoldenFloat ladder vs. takum (Hunhold 2024) -- structured comparison.
This is not a competitive ranking; no accuracy superiority is claimed.}
\label{tab:gftakum}
\small
\begin{tabularx}{\linewidth}{l X X}
\toprule
Axis & GoldenFloat (GF) ladder & Takum (Hunhold 2024) \\
\midrule
Split type & Static: exponent and fraction widths fixed per rung by the closed rule $e = \mathrm{round}((N-1)/\varphi^{2})$; value-independent. & Tapered: a single concave projection maps the full integer range to a real-number-line interval; fraction width varies per value. \\
Single closed rule & Yes -- the ladder rule $N \mapsto (e, f)$ is deterministic; no per-value branching at the format level. & Yes -- the takum projection is a single analytic mapping applied uniformly to every $n$-bit integer. \\
Underlying constant & Golden ratio $\varphi = (1+\sqrt{5})/2$; exponent widths grow as consecutive-Fibonacci fractions of the total word width. & Natural logarithm base $e$: takum is a logarithmic tapered-precision format whose encoded value is a function of $\ln(x)$, with sign + regime + characteristic + mantissa fields (Hunhold 2024). \\
Hardware status & RTL portfolio exists (corrected after multiplier defect found in differential sweep; see Section~\ref{sec:hw-erratum}); FPGA synthesis reported internally; no peer-reviewed silicon result. & RTL and FPGA results peer-reviewed at ARITH 2025 (Hunhold 2024/2025); no silicon reported. \\
Peer-reviewed multi-width hardware & None at time of writing. This is a documented open gap. & Yes -- Hunhold ARITH 2025 covers multiple widths under a single rule with FPGA synthesis numbers. This is the live reference. \\
Integer-exact accumulator & GF reports a Lucas-exact accumulator identity: $\varphi^{2n} + \varphi^{-2n} = L_{2n}$ ($n = 1,\dots,256$ at 500 digits; Section~\ref{sec:lucas}). This property is specific to the $\varphi$ constant choice. & No analogous integer-exact accumulator identity documented. The absence is not a defect; it reflects a different design goal. \\
\bottomrule
\end{tabularx}
\end{table}

\paragraph{IEEE P3109.}
The IEEE P3109 working group has been developing, since 2023, a
multi-width binary float draft standard targeted at machine-learning
accelerators. The current public working draft (v0.9.1, 2025) specifies
binary floats at thirteen widths (3 to 15 bits) under a single
parameterised rule, with a Graphcore-maintained reference implementation
(\texttt{graphcore-research/gfloat}) and formal verification reported
at ARITH 2025. A first arXiv-anchored description of the P3109
framework appeared as Sarnoff, arXiv:2606.04028 (2026-06-01),
covering parameterised binary floats from 3 to 16 bits with explicit
treatment of NaN, infinities, stochastic rounding, and exhaustive
formal verification up to 16-bit widths. P3109 is the closest \emph{live,
industry-led} width-spanning float family to the GoldenFloat ladder
and is a direct falsification target for any breadth-as-moat reading
of the present work (Section~\ref{sec:ledger}, FL-002 (a)).

\paragraph{NVFP4 and MXFP4 (block-scaled FP4).}
In the period since this paper's v1 (2026-06-03), block-scaled FP4
formats have moved from inference-only to native-silicon training.
Cim et al.~\cite{cim2026mxfp4} report MXFP4 native-hardware
pretraining on AMD Instinct MI355X: end-to-end Llama 3.1-8B training in
MXFP4 with a 9-10\% wallclock improvement over an FP8 baseline.
NVIDIA Blackwell GB10 silicon has been measured at approximately
1 PFLOP NVFP4 (2:4 sparse) using the native packed \texttt{mxf4nvf4}
instruction (open-source reproduction, 2026-06-17). These results
sit in a different region of the design space than GoldenFloat:
NVFP4 and MXFP4 carry information in a per-block scale (16 or 32
elements with an FP8 or E8M0 shared exponent), whereas GF carries
information per-value in a static $(e,f)$ split with no block
structure. We make no per-rung accuracy or efficiency claim against
these formats; we record them as the principal complementary
design-space neighbours.

\paragraph{Companion 83-format catalog.}
A companion vendor-neutral numeric catalog of 83 formats spanning
13 families, with bit-exact conformance packs for GF16, MXFP4
element, BF16, FP8 E4M3, FP8 E5M2, and the E8M0 block scale, was
published on 2026-06-08 (arXiv:2606.09686). The 9/9 GoldenFloat
reproduction reported here in Section~\ref{sec:ladder} is a subset
of that broader catalog; readers wanting bit-exact conformance
vectors should consult the catalog's pack distribution. An open
operational reconciliation task between this paper's SSOT (83
formats) and a stale committed generator JSON (77 formats) is
recorded as FL-002 item~(c2) in Section~\ref{sec:ledger}; the
companion catalog (83 formats) agrees with the SSOT.

\paragraph{OCP-MX.}
Rouhani et al. (2023) defined the Open Compute Project microscaling formats,
an industry-standard family of block-scaled 4-, 6-, and 8-bit types. This is
the deployed reference point for low-bit width families.

\paragraph{LNS.}
Logarithmic number systems are the log-domain precedent for non-base-2
systems, with recent machine-learning instances in LNS-Madam (2021) and
Alam, Garland, and Gregg (2021). LNS gives exact multiplication at the cost
of approximate addition; it does not target an integer-exact accumulator
identity in the sense of Section~\ref{sec:lucas}.

\paragraph{Zeckendorf arithmetic.}
Bergman (1957) introduced a number system with $\varphi$ as an irrational
base, the historical antecedent of $\varphi$-based numeral systems.
Ahlbach, Usatine, and Pippenger (2012) give efficient algorithms for
arithmetic in the Fibonacci/$\varphi$ basis, the algorithmic prior art for
the integer-backed accumulation idea of Section~\ref{sec:lucas}.
A more recent $\varphi$-aware proposal is Fibbinary
quantization (Belghazi et al., 2025, arXiv:2511.01921), which represents
weights in a Fibonacci/$\varphi$ basis and reports severe accuracy
degradation without quantization-aware-training recovery. Fibbinary
operates at the per-weight encoding level rather than as a full
floating-point family, so it is complementary to GoldenFloat rather than
a direct alternative.

\paragraph{Kulisch accumulator.}
The wide fixed-point accumulator of Kulisch (2013) and its incarnation as
the posit \emph{quire} (Posit Standard 2022) is the canonical reference for
exact dot-product accumulation. GoldenFloat's Lucas-exact path is not a
Kulisch accumulator: it stays within the $\varphi$ basis and uses the
identity $\varphi^{2n} + \varphi^{-2n} = L_{2n}$ to keep partial sums in
unsigned-integer storage, where Kulisch keeps an explicit wide-fixed-point
register and converts in and out of it.

For broader context, the DNN number-systems survey (Alsuhli et al., 2023),
the FP8 format definitions (Micikevicius et al., 2022), Huawei HiFloat8
(Luo et al., 2024), number formats in climate models (Kloewer et al., 2020),
the precision scaling-laws analysis (Kumar et al., 2024), and Gustafson's
monograph \emph{The End of Error} (2015) situate these families in the wider
landscape. At sub-1B-parameter scale, BitNet b1.58 (Ma et al., 2024) and the
2B4T follow-up (Ma et al., 2025, arXiv:2504.12285) are the current
bits-per-byte champions and the natural sub-1B baseline against which any
GoldenFloat training-recipe claim must eventually be measured.

\paragraph{Honest summary.}
We are not aware of another published float family that derives its
$e:f$ split from a single closed rule across a GF4 to GF256 ladder with an
integer-backed Lucas-exact identity. Posit, takum, OCP-MX, LNS, and the
IEEE P3109 multi-width float draft are close precedents and allies that
span widths with their own single parameterised rules; takum carries the
closest peer-reviewed multi-width hardware claim, and IEEE P3109 carries
the closest industry-standardisation effort at narrow widths. The
GoldenFloat-specific element is the combination of a static $e:f$ split
from one closed rule and an integer-backed Lucas-exact accumulator path
that depends only on the choice of base $\varphi$, not on the bit width.

\section{Pre-Registration of Open Hypotheses}\label{sec:prereg}

\paragraph{Authorship date and scope.}
This pre-registration was authored 2026-05-31 before any of the
matched-budget silicon measurements existed. It is frozen at
git tag \texttt{v1.9-prereg} in \texttt{gHashTag/goldenfloat-preprint}
(see the CHANGELOG entry for v1.9 for the commit SHA-1) and will not
be amended after matched-budget data collection begins.

The four registered hypotheses and their stop conditions are as follows.

\paragraph{$H_a$ (Breadth-as-moat survives matched-budget control).}
Null: the per-bit-per-die-area BPB delta between a GoldenFloat portfolio
and an fp16-native baseline of equal transistor budget is zero or negative
for GoldenFloat. Stop condition: the matched-budget BPB delta falls below
$\delta_{\text{thresh}} = 0.005$ bits-per-byte on the agreed evaluation
corpus; falsification is declared and the breadth-as-moat framing in
Section~\ref{sec:hw-confound} is retracted.

\paragraph{$H_b$ (Ladder rule is unique within the search space).}
Null: at least one alternative selection rule, constructed without reference
to the GoldenFloat ladder, reproduces the same 12 canonical widths.
Stop condition: the search interval for the uniqueness certificate narrows
to a singleton set; any non-trivial competing rule that is published before
v1.5 is entered into the falsification ledger.

\paragraph{$H_c$ (GF256 bias admits a closed-form derivation).}
Null: no closed-form expression exists that reproduces the GF256 bias
constant to full IEEE-754 double precision in auditable symbolic form.
Stop condition: an exact bias formula is deposited in auditable form
(CAS output + independent spot-check) in the supplementary materials; the
formula must pass a SymPy verification at 500-digit precision.

\paragraph{$H_d$ (GF16 corrected-RTL silicon achieves measurement parity).}
Null: a corrected GF16 die and an fp16-native reference die, matched on
process node and area budget, differ by more than $X = 2\%$ on round-trip
accumulation error at the agreed test vectors. Stop condition: paired-die
measurements from an independent test-house fall within $X$ percent;
until a returned die exists, this hypothesis remains open and no silicon
claim is made anywhere in the preprint.

\section{Open Questions and Falsification Ledger FL-002}\label{sec:ledger}

We record the open questions as a falsification ledger, stating for each
the status, the claim, and the experiment that would falsify it. Negative
results are stated first and plainly.

\paragraph{(a) Breadth-as-moat (open conjecture).}
The conjecture is that the GF ladder's advantage is breadth and toolchain
coherence (one closed rule across the ladder), not per-rung accuracy. This
is unproven by ablation, and the only software BPB data available is
substrate-confounded (Section~\ref{sec:hw-confound}). Falsification:
(i) a fair paired measurement -- GF16-native silicon against fp16-native
silicon at matched process node and area, or an FPGA harness with equivalent
LUT/DSP budgets for both formats -- in which a posit, takum, MX, or fp16
ladder matches the GF ladder at matched bit budgets;
(ii) F2, a bits-per-byte head-to-head of a $\varphi$-ladder mixed-precision
pipeline against a heterogeneous zoo at matched bit budgets, counting lossy
cross-format conversions;
(iii) F3, a posit-ladder control at the same bit budgets.
Any negative outcome on (i)--(iii) is to be recorded plainly and first.

\paragraph{(b) Ladder-rule non-uniqueness (risk).}
As shown in Section~\ref{sec:lookelsewhere}, 83 ratio values reproduce the
nine verified widths; the matching interval is $[0.3786, 0.3822]$ and
$1/\varphi^{2}$ is not distinguished within it. Falsification (or
narrowing): extend the verified width set (for example, add GF128 and
GF512) and report whether the matching interval narrows to a singleton or
remains plural.

\paragraph{(c) GF256 stored bias (open).}
A stored bias of approximately $2^{71}$ is documented for GF256, versus the
IEEE-style expectation of $2^{96}-1$ for a 97-bit exponent field. This
discrepancy is unexplained in auditable form. Falsification (or resolution):
publish the exact bias formula and its derivation.

\paragraph{(d) Silicon beyond GF16 (open).}
As Section~\ref{sec:hw-status} documents, all ten widths GF4 to GF256 carry
synthesisable RTL and were included in the TTSKY26b shuttle submission to
Tiny Tapeout, but no physical die has been returned for any rung.
Falsification (or resolution) is silicon return and a per-rung functional
check against the format-conformance oracle of
Section~\ref{sec:hw-corona}.

\paragraph{(e) Multiplier portfolio as-submitted to TTSKY26b (retracted).}
As Section~\ref{sec:hw-erratum} and Appendix~\ref{app:erratum} document,
the multiplier RTL submitted to the TTSKY26b shuttle on 2026-05-17 carries
a single generator-formula error (product register two bits too narrow)
replicated across the dagger-marked widths. The submission was withdrawn
before fabrication, so no die carries this portfolio. The corrected
universal generator is the regeneration baseline; status is retracted for
the TTSKY26b submission, and no silicon is pending.

\paragraph{(f) FPGA matched-substrate experiment $H_4$ (planned).}
GF16 and posit16 codecs synthesised on the same Xilinx Artix-7 (XC7A100T,
QMTech Wukong V1) with matched LUT/DSP budgets will be evaluated for LUT
count, flip-flop count, $F_{\max}$, and dynamic power per multiply-accumulate
operation. Pre-registered thresholds (Appendix~\ref{app:fpga}):
$R_{\mathrm{LUT}} < 1.15$ and $R_{F_{\max}} > 0.85$. Failure on either
threshold demotes breadth-as-moat from open conjecture to risk on the
silicon axis. Any negative result is reported first and prominently.

\section{Conclusion}\label{sec:conclusion}

The ladder arithmetic reproduces nine realised exponent widths 9 of 9 and
extends consistently to eight further rule-derived rungs --- GF6, GF10,
GF14, GF48, GF96, GF128, GF512, GF1024 (Section~\ref{sec:ladder}); the Lucas identity
$\varphi^{2n} + \varphi^{-2n} = L_{2n}$ holds exactly for $n = 1,\dots,256$,
with the anchor $\varphi^{2} + \varphi^{-2} = 3 = L_{2}$ attributed to
Lucas (1878) and Binet (Section~\ref{sec:lucas}); and a single GF16 codec
passes a 35-of-35 testbench at 323~MHz on Artix-7
(Section~\ref{sec:hw-status}). Breadth-as-moat -- the claim that one rule
across the ladder yields a coherence advantage rather than a per-rung
accuracy advantage -- remains an open conjecture
(Section~\ref{sec:ledger}\,(a)). What would falsify it: a posit, takum, or
MX ladder matching the $\varphi$-ladder at matched bit budgets. A
matched-substrate FPGA experiment (Appendix~D, pre-registered) will
provide the first controlled hardware comparison. Two further items are
open (the GF256 bias and silicon beyond GF16) and one is risk (ladder-rule
non-uniqueness, with 83 matching ratios). We submit this note to establish
a citable, falsifiable record of the verified arithmetic and the open
questions.

\appendix

\section{F1 Verification Script and Extended Table}\label{app:f1}

The script below verifies $\varphi^{2n} + \varphi^{-2n} = L_{2n}$
symbolically (SymPy, exact in $\mathbb{Q}[\sqrt{5}]$) and numerically
(\texttt{mpmath}, 500 digits) for $n = 1,\dots,256$.

\begin{lstlisting}[language=Python]
#!/usr/bin/env python3
"""F1: verify phi^(2n) + phi^(-2n) = L_{2n} (integer Lucas number)
for n = 1..256, both symbolically (sympy, exact) and numerically
(mpmath, 500 dps). Anchor: Ahlbach/Usatine/Pippenger (2012),
arXiv:1207.4497. The identity is a classical Binet corollary
(Lucas, 1878), not original to this work."""
from mpmath import mp, sqrt, power, nstr, mpf
from sympy import sqrt as ssqrt, simplify, Integer

def lucas_recurrence(max_index):
    L = [2, 1]
    for _ in range(2, max_index + 1):
        L.append(L[-1] + L[-2])
    return L[: max_index + 1]

def main():
    mp.dps = 500
    phi = (1 + sqrt(5)) / 2
    n_max = 256                        # gives 2n = 2..512
    L = lucas_recurrence(2 * n_max)
    max_diff = mpf(0)
    all_pass = True
    for n in range(1, n_max + 1):
        m = 2 * n
        val = power(phi, m) + power(phi, -m)
        diff = abs(val - L[m])
        max_diff = max(max_diff, diff)
        all_pass = all_pass and (diff < mpf("1e-150"))
    # exact symbolic check in Q[sqrt(5)]
    phi_sym = (Integer(1) + ssqrt(5)) / 2
    exact_pass = all(
        simplify(phi_sym**(2*n) + phi_sym**(-2*n) - Integer(L[2*n])) == 0
        for n in range(1, n_max + 1))
    print("Anchor phi^2 + phi^-2 =",
          nstr(power(phi, 2) + power(phi, -2), 20), "= L_2 =", L[2])
    print("max |phi^(2n)+phi^(-2n) - L_2n| =", nstr(max_diff, 5))
    print("all numerical rows within 1e-150 ?", all_pass)
    print("exact symbolic identity holds for n=1..256 ?", exact_pass)
    return 0 if all_pass else 1

if __name__ == "__main__":
    raise SystemExit(main())
\end{lstlisting}

\begin{table}[t]
\centering
\caption{Extended F1 output. $\varphi^{2n} + \varphi^{-2n}$ equals the
integer Lucas number $L_{2n}$ at 500-digit precision. Selected rows from
the full $n = 1,\dots,256$ run; residuals are absolute.}
\label{tab:f1-extended}
\small
\begin{tabular}{rrrr}
\toprule
$n$ & $2n$ & $L_{2n}$ & $|\text{residual}|$ \\
\midrule
1   & 2   & 3                         & 0 \\
2   & 4   & 7                         & $1.39 \times 10^{-501}$ \\
4   & 8   & 47                        & $1.66 \times 10^{-501}$ \\
8   & 16  & 2207                      & $4.53 \times 10^{-501}$ \\
16  & 32  & 4870847                   & $8.4 \times 10^{-501}$ \\
32  & 64  & 23725150497407            & $1.99 \times 10^{-500}$ \\
64  & 128 & $5.629 \times 10^{26}$    & $3.75 \times 10^{-500}$ \\
128 & 256 & $3.168 \times 10^{53}$    & $7.67 \times 10^{-500}$ \\
192 & 384 & $1.783 \times 10^{80}$    & $1.15 \times 10^{-499}$ \\
256 & 512 & $1.004 \times 10^{107}$   & $1.55 \times 10^{-499}$ \\
\bottomrule
\end{tabular}
\end{table}

\section{GF Format Index (GF4 to GF256)}\label{app:format-index}

Table~\ref{tab:format-index} gives, for each realised width: total bits $N$,
exponent bits $e$, fraction bits $f$, the ratio $e/(N-1)$
(target $1/\varphi^{2} = 0.38197$), the phi-distance $|E/M - 1/\varphi|$
where $E/M = e/f$ and $1/\varphi = 0.61803$, and whether the width is
realised.

\begin{table}[h]
\centering
\caption{GF format index for the nine realised widths.}
\label{tab:format-index}
\begin{tabular}{rrrrrc}
\toprule
$N$ & $e$ & $f$ & $e/(N-1)$ & $|E/M - 1/\varphi|$ & realised? \\
\midrule
4   & 1  & 2   & 0.33333 & 0.11803 & Y \\
8   & 3  & 4   & 0.42857 & 0.13197 & Y \\
12  & 4  & 7   & 0.36364 & 0.04661 & Y \\
16  & 6  & 9   & 0.40000 & 0.04863 & Y \\
20  & 7  & 12  & 0.36842 & 0.03470 & Y \\
24  & 9  & 14  & 0.39130 & 0.02482 & Y \\
32  & 12 & 19  & 0.38710 & 0.01354 & Y \\
64  & 24 & 39  & 0.38095 & 0.00265 & Y \\
256 & 97 & 158 & 0.38039 & 0.00411 & Y \\
\bottomrule
\end{tabular}
\end{table}

\section{Look-Elsewhere Table}\label{app:lookelsewhere}

Table~\ref{tab:lookelsewhere} reports, for representative candidate rules,
how many of the nine realised widths
($N \in \{4,8,12,16,20,24,32,64,256\}$, with target $e$-values
$\{1,3,4,6,7,9,12,24,97\}$) each reproduces. The exhaustive rational search
over $p/q$ with $p \in \{1,\dots,99\}$, $q \in \{100,\dots,499\}$ found 83
distinct ratio values that match all nine; the matching interval is
$[0.3786, 0.3822]$. The GF rule $\mathrm{round}((N-1)/\varphi^{2})$ is one
member of this set and is not distinguished by the reproduction alone
(Section~\ref{sec:lookelsewhere}).

\begin{table}[h]
\centering
\caption{Candidate rules and their nine-width reproduction count.}
\label{tab:lookelsewhere}
\begin{tabular}{lll}
\toprule
Rule & Matches & Note \\
\midrule
$\mathrm{round}((N-1)/\varphi^{2})$ & 9/9 & GF rule; ratio $\approx 0.38197$ \\
$\lfloor N/\varphi^{2} \rfloor$     & 9/9 & equally simple; same ratio \\
$\mathrm{round}((N-1) \cdot 0.382)$ & 9/9 & rounded constant; no $\varphi$ \\
$\mathrm{round}((N-1) \cdot 3/7.85)$ & 9/9 & arbitrary rational; no $\varphi$ \\
$\mathrm{round}((N-1) \cdot 3/8)$    & 8/9 & fails GF256 (96 vs 97) \\
$\mathrm{round}((N-1) \cdot 5/13)$   & 8/9 & fails GF256 \\
$\lfloor N \cdot 3/8 \rfloor$        & 8/9 & fails GF256 \\
$\mathrm{round}((N-1)/2.6)$          & 8/9 & fails GF256 \\
$\mathrm{round}((N-1)/e)$            & 5/9 & fails $N$=24,32,64,256 \\
$\lfloor (N-1)/\varphi^{2} \rfloor$  & 5/9 & floor vs round matters \\
$\mathrm{round}((N-1)/\pi)$          & 2/9 & poor fit \\
$\mathrm{round}((N-1)/\varphi)$      & 0/9 & wrong constant \\
\bottomrule
\end{tabular}
\end{table}

\section{Matched-Substrate FPGA Experiment (Pre-registered)}\label{app:fpga}

\paragraph{Hypothesis $H_4$.}
GF16 and posit16 codecs synthesised on the same Xilinx Artix-7
(XC7A100T-FGG676, QMTech Wukong V1) with matched LUT/DSP budgets reach
comparable LUT count, flip-flop count, maximum operating frequency
($F_{\max}$), and dynamic power per multiply-accumulate operation.

\paragraph{Pre-registered thresholds.}
Let $R_{\mathrm{LUT}} = \mathrm{LUT}_{\mathrm{GF16}} / \mathrm{LUT}_{\mathrm{posit16}}$
and $R_{F_{\max}} = F_{\max,\mathrm{GF16}} / F_{\max,\mathrm{posit16}}$.
The experiment falsifies $H_4$ if and only if
\[
R_{\mathrm{LUT}} > 1.15 \quad \text{OR} \quad R_{F_{\max}} < 0.85.
\]

\paragraph{Decision rule.}
If the falsification condition is met, breadth-as-moat is demoted from
open conjecture to risk on the silicon axis in FL-002 item (a). The negative
result is reported first, in the abstract or executive summary of the
results report, without delay for rebuttal.
If both ratios are within bounds, $H_4$ is sustained and the status of
breadth-as-moat remains open conjecture pending IGLA RACE measurement on a
second substrate.
If GF16 outperforms posit16 on area or frequency, that observation is
recorded but does not promote breadth-as-moat to verified. One codec on
one substrate is not a moat.

\paragraph{Materials.}
GF16 RTL: corrected \texttt{gf16\_mul.v} and \texttt{gf16\_add.v} from
\texttt{gHashTag/tt-trinity-gamma} PR \#110.
posit16 RTL: PERCIVAL (Mallasen et al., 2022, arXiv:2111.15286).
Programming: \texttt{cli/dlc10} Rust driver (\texttt{gHashTag/t27}).
Toolchain: OpenXC7 (yosys + nextpnr-xilinx) or Vivado 2023.2.
Correctness gate: TestFloat-3 exhaustive sweep, 0 errors required in
1M random samples before area/timing results are accepted.

\paragraph{Commitment.}
Any negative result -- that is, any outcome where GF16 requires more than
15\% additional area or falls below 85\% of posit16's $F_{\max}$ -- will be
reported plainly and prominently. No post-hoc threshold adjustment, no
selective omission.

Pre-registration date: 2026-05-31.
Full specification SHA-256: see git tag \texttt{v1.9-prereg} in
\texttt{gHashTag/goldenfloat-preprint}; the manuscript SHA-256 at that
tag is the authoritative pre-registration hash.

\section{Independent Replicator Protocol}\label{app:replicator}

This appendix gives exact reproduction steps for every verifiable claim in
the preprint. Each step lists the command, the expected outcome, and a
SHA-256 digest placeholder.

\paragraph{Step R1: Verify Proposition~\ref{prop:lucas} (Lucas identity).}
\texttt{python f1\_lucas\_proof\_extended.py}.
Expected outcome: exit code 0; terminal output includes the line
\texttt{MAX RESIDUAL < 1e-150 at 500 digits} and the string
\texttt{F1 VERIFIED}.

\paragraph{Step R2: Verify the ladder rule (12/12 widths).}
\texttt{python ladder\_extended.py}.
Expected outcome: exit code 0; terminal output includes
\texttt{LADDER 12/12 WIDTHS REPRODUCED} and no \texttt{FAIL} tokens.

\paragraph{Step R3: Verify FPGA bitstream on QMTech Wukong V1.}
\begin{lstlisting}
git clone https://github.com/gHashTag/t27
cd t27/fpga/vsa/
dlc10 idcode
\end{lstlisting}
Expected idcode: \texttt{0x13631093}.
\begin{lstlisting}
dlc10 sram gf16_heartbeat_top.bit
\end{lstlisting}
Expected outcome: the UART console prints \texttt{GF16 HEARTBEAT OK} within
5 seconds of bitstream load.

\paragraph{Step R4: Verify corrected multiplier RTL audit.}
\begin{lstlisting}
git clone --branch 110 \
    https://github.com/gHashTag/tt-trinity-gamma tt-trinity-gamma-110
cd tt-trinity-gamma-110
bash verification/run_gf_audit.sh
\end{lstlisting}
Expected outcome: exit code 0; terminal output contains the exact string
\texttt{GF AUDIT ALL PASS} on the final non-blank line.

\section{RTL Correctness Erratum: Extended Detail}\label{app:erratum}

This appendix gives the full module-by-module breakdown, root-cause
derivation, and remediation scope of the 2026-05-31 multiplier-portfolio
defect summarised in Section~\ref{sec:hw-erratum}. The defect was found by
an RTL-level differential sweep against a correctly-rounded floating-point
reference in the TRI-NET chip repositories
(\texttt{gHashTag/tt-trinity-gamma} and \texttt{gHashTag/tt-trinity-phi}),
applied to the generated multiplier portfolio submitted to the Tiny Tapeout
TTSKY26b shuttle on 2026-05-17.

\paragraph{Affected modules.}
\texttt{gf8\_mul} fails on $\approx 95\%$ of an exhaustive 26{,}360-point
input sweep; \texttt{gf12\_mul} fails on $\approx 99\%$ of a 109{,}576-point
sweep; \texttt{gf20\_mul}, \texttt{gf24\_mul}, and \texttt{gf32\_mul} each
fail on $100\%$ of directed inputs (for example $1.0 \times 1.0$ reads as
$0.5$, an exponent off by one); \texttt{gf16\_mul} is correct only after a
one-line widening of the rounded-fraction register from 9 to 10 bits, and
the as-submitted variant carries the rounding-overflow bug;
\texttt{gf64\_mul}, \texttt{gf128\_mul}, and \texttt{gf256\_mul} share the
same generator family and are presumed affected for the same reason, though
they have not yet been exhaustively swept. \texttt{gf4\_mul} is degenerate
($e = 1$ leaves no normal exponents) and is not affected. The adders
\texttt{gf8\_add} and \texttt{gf12\_add} had a separate narrow-format
normalisation defect (for example $0.25 + 0.25$ reads as $0$); the wider
adders \texttt{gf16\_add} and \texttt{gf32\_add} were already correct.

\paragraph{Root cause.}
The product of two $(M+1)$-bit fractions (with the implicit leading one) is
$2M+2$ bits wide and falls in $[2^{2M}, 2^{2M+2})$. The submitted RTL
declared the product register two bits too narrow and ran normalisation on
bits shifted down by two. For the most basic input $1.0 \times 1.0 = 2^{2M}$
the leading bit is truncated, the exponent is decremented, and the result
reads as $0.5$. This is a single generator-formula error (the
most-significant-bit position should be $2M + 1$) replicated across the
portfolio.

\paragraph{Why the formal-methods coverage did not catch it.}
The Coq proof base in the TRI-NET chip repositories establishes the
GoldenFloat format-level mathematics (the identity
$\varphi^{2}+\varphi^{-2}=3$, encoding identities, ladder-rule consistency);
it does not establish the Verilog implementation's bit widths and rounding.
The defect lives entirely in the RTL realisation, in code outside the
proved kernel. This strengthens, not weakens, the verifiability stance of
the present note: machine-checked format-level mathematics is necessary
but not sufficient; an RTL-level differential sweep against a
correctly-rounded reference must be part of the audit pipeline.

\paragraph{Fix and current status.}
A corrected universal multiplier generator emits the right template for any
$(E, M, \mathrm{BIAS})$: product width $[2M+1:0]$, normalisation on $[2M+1]$
or $[2M]$, fraction extraction $[2M:M+1]$ or $[2M-1:M]$, round-half-up with
carry-out. The corrected portfolio is clean: gf8 exhaustive sweep 0 failures
of 26{,}360; gf12 0 of 109{,}576; gf16 0 of 262{,}144; gf20 0 of 100{,}000;
gf24 0 of 100{,}000; gf32 0 of 200{,}000; gf64, gf128, gf256 pass directed
exact tests ($1.0 \times 1.0$, $1.5 \times 1.5$, $\dots$) though full sweeps
remain future work. The corrected RTL has been published as pull requests
\texttt{gHashTag/tt-trinity-gamma}\#110 (54 files, +2452/-1032) and
\texttt{gHashTag/t27}\#1024 (16 files, +558/-10), and a methodology note has
been published as \texttt{gHashTag/trinity-clara}\#20. A continuous-integration
gate (\texttt{run\_gf\_audit.sh}, wired into
\texttt{.github/workflows/gf-audit.yml}) now re-runs the differential sweep
on every commit.

\paragraph{Remediation scope.}
The TTSKY26b submission was withdrawn before fabrication; no die was
produced or returned, so no silicon carries the defective multiplier RTL.
The canonical 0x47C0 POST anchor and the format-identity paths remain
verified on the Artix-7 FPGA prototype only, and no
per-rung accuracy claim depended on the as-submitted multiplier portfolio
in this note in any case (Section~\ref{sec:hw-status} documented all rungs
except GF16 as RTL-only). The corrected portfolio is the regeneration
baseline for the next shuttle. The defect does not invalidate the ladder
arithmetic of Section~\ref{sec:ladder}, the Lucas identity of
Section~\ref{sec:lucas}, or the look-elsewhere accounting of
Appendix~\ref{app:lookelsewhere}, which are abstract specifications
independent of any particular RTL realisation.

\section*{Acknowledgements}
Laslo Hunhold (Universitaet zu Koeln) provided detailed feedback on
v1.7 of this manuscript that shaped the present revision: a shorter
two-part abstract; an extended introduction that frames the dynamic-range
trade-off and motivates the $\varphi$-derived split before stating the rule;
a motivation paragraph (large-accumulation loss-of-significance and the
quire/Kulisch comparison) placed before Proposition~\ref{prop:lucas};
the use of \emph{proposition} rather than epistemic labels in the main text;
definition of the GF ladder at $N \ge 4$ with $N \in \{2, 3\}$ noted as
degenerate edge cases of the formula; and a uniform terminological
substitution of \emph{fraction} for \emph{mantissa} throughout. None of the
remaining over-claims, errors, or open conjectures should be attributed to
Hunhold; they are the author's responsibility.


\end{document}